\documentstyle[twocolumn,aps]{revtex}

\begin{document}
\title{\bf Convective Nonlinearity in Non-Newtonian Fluids}
\author{
{\it Hubert Temmen,$^{1}$ Harald Pleiner,$^{2}$
 Mario Liu$^{3}$  and Helmut R. Brand$^{4}$}\\
 [0.3cm]
{\small  1. EMA, Universit\"at der Bundeswehr
Hamburg, Holstenhofweg 85, 22043 Hamburg \\ 2.
Max-Planck-Institut f\"ur Polymerforschung,
 Postfach 3148, 55021 Mainz \\
 3. Institut f\"ur Theoretische Physik,
Universit\"at Hannover, 30167 Hannover \\
 4. Theoretische Physik III, Universit\"at Bayreuth,
 95440 Bayreuth, Germany}}
\date{PRL  84, 3223, (2000)}
\maketitle
\abstract{In the limit of infinite yield time for
stresses, the hydrodynamic equations for viscoelastic,
Non-Newtonian liquids such as polymer melts must reduce
to that for solids. This piece of information suffices to
uniquely determine the nonlinear convective derivative,
an ongoing point of contention in the rheology
literature.}}
\draft{}
Hydrodynamics, the macroscopic description of condensed
systems in the low frequency, long wavelength
limit~\cite{mpp,forster}, is a well-established,
systematic approach in condensed matter physics. This
method has been applied over the past decades to many
systems, including simple fluids \cite{llhydro}, various
liquid crystals~\cite{mpp,lc}, and the superfluid phases
of $^3$He and $^4$He~\cite{ml}.

All viscoelastic non-Newtonian fluids behave as Newtonian
ones at low frequencies, and as solids at higher
frequencies. A consistent hydrodynamic description needs
to reflect this fact and must therefore contain, as
special cases, both the hydrodynamic theory for isotropic
liquids and solids. The liquid limit is well heeded in
the polymer literature and universally correctly
implemented \cite{truesdell,noll}. The solid limit is
problematic, as we shall see, and compatibility
especially in the nonlinear regime of large displacements
and rotations has so far proven elusive. The reason
behind it is probably the lack of a consistent
hydrodynamic theory for solids.

The last sentence may come as a surprise, but the point
we are making here is: Although both the nonlinear
elasticity theory~\cite{truesdell} and the linearized
hydrodynamics for crystals~\cite{mpp,llelastic,cohen} are
well known and established, a consistent hydrodynamic
theory that includes both nonlinear and irreversible
terms is not -- in spite of some insightful
papers~\cite{pl}. One of the obstacles is that such a
theory necessarily employs a strain tensor different from
the one customarily used~\cite{llelastic}. The usual
strain tensor is of the Lagrange type, derived from
equations of motion for mass points, while a framework to
set up hydrodynamic equations including dissipative terms
only exists in the Eulerian description -- which
considers evolution of field variables at spatial points.
Consistency forbids a mixing of both descriptions and
requires an Eulerian strain tensor~\cite{Lub2}. (We note
that the linear hydrodynamic theory may mix both
descriptions, as the smallness of the displacements
ensures that the discrepancy is negligible.)

The presentation of the nonlinear hydrodynamic theory for
solids is what we shall do first. Then these equations
are generalized for non-Newtonian fluids by adding
relaxation-type terms to account for a finite yield time
of the stresses, such that in the high frequency limit
the theory is unchanged, but in the low frequency limit
only the terms of the isotropic liquid hydrodynamics
remain. So, by ensuring the valid liquid and solid
limits, this approach leads to the correct hydrodynamic
theory for any liquids displaying viscoelasticity. It is
of great importance for rheology, as many competing
theories exist, which differ especially in their
respective nonlinear convective derivatives. All are
employed in the literature, with scant hope for
consensus.

Comparing our result with the literature, we find that
none of the convective nonlinearities suggested in the
rheology literature \cite{truesdell,noll} contains the
correct solid-limit, putting justified doubts on their
validity. (We do obtain, for the Eulerian strain and in
the limit in which it is small, the so called ``upper
convected derivative".)

Note that the insistence on the solid limit also
determines the choice of variables, being that of a
solid: the strain and the conserved quantities. As the
solid is the more complicated of the two limits, there is
no reason to, in addition, take the stress as an
independent variable, as most theories in the rheology
literature do, see for instance Chap.~7-9 in the first
of~\cite{noll}. Aside from unnecessarily making a derived
quantity independent, this approach also commits a
cardinal hydrodynamic sin, because the stress as a
hydrodynamic flux contains reactive and dissipative
parts, and does not possess a well-defined time reversal
parity -- without which we have no way of applying the
Onsager relations.

We now introduce the nonlinear hydrodynamic theory
of solids. A proper description relies on two
coordinates: the actual spatial coordinate $r_i$,
specifying a point in an elastic body, and the
coordinate $a_i$ this point possesses in the absence
of any stresses. More carefully, starting from a
stress-free elastic body, we consider a point with
the initial coordinate $a_i$. As the body is
translated, rotated, compressed and sheared, this
point is displaced to $r_i$ -- especially in soft
matter generally rather remote from $a_i$. Since all
points of the body have a unique pair of $a_i$ and
$r_i$, the function $r_i(a_m)$ is unique and
invertible, the result of which is denoted as
$a_i(r_m)$. For briefness, we shall refer to all
$r_i$ as the real space, and to all $a_i$ as the
initial space.

As discussed in most books on elasticity theory, see
eg~\cite{llelastic}, the elastic energy depends on the
change in the distance between any two neighbouring
points, from ${\rm d}a^2_i$ to ${\rm d}r^2_i$. Defining
the displacement vector as $u_i(a_m)=r_i(a_m)-a_i$, and
the strain tensor as $U^L_{ik}= \textstyle \frac{1}{2}
[\partial u_i/\partial a_k+ \partial u_k/\partial a_i+
(\partial u_j/\partial a_i)\cdot(\partial u_j/\partial
a_k)]$, we have ${\rm d}r^2_i(a_m)-{\rm
d}a^2_i=2U^L_{ik}{\rm d}a_i{\rm d} a_k$, and know to take
the energy density as a function of $U^L_{ik}$, to lowest
order simply as $\epsilon= \textstyle\frac{1}{2} K_{ikjm}
U^L_{ik}U^L_{jm}$. The important point here is that both
the strain tensor and the energy density are functions of
the initial coordinate $a_m$, a notation that we shall
refer to as Lagrangian -- hence the superscript in
$U^L_{ik}$.

Contrast this with the energy density of an isotropic
liquid in its rest frame, a function of the mass and
entropy density, $\epsilon(\rho, s)$ -- or equivalently,
${\rm d}\epsilon=T{\rm d}s+\mu{\rm d}\rho$. All
variables, including the conjugate ones, temperature $T$
and chemical potential $\mu$, are here functions of the
real coordinate $r_m$. As a result, the spatial
dependence of (say) the temperature is quite independent
of the liquid's compressional state. This is the Euler
notation, and its basic advantage is that physics, which
we insist must be local, is also expressed in local
terms, accounted for by quantities at the real
coordinates $r_m$. Consider for instance the diffusive
heat current, which is given by the local gradient of the
temperature, $\sim\partial T(r_m)/\partial r_k$, only in
the Eulerian description.

Returning to solids, we have two choices: First, take all
variables including especially the temperature and
chemical potential as functions of $a_m$, and employ them
with the strain tensor $U^L_{ik}$. This would be
consistent, but highly inconvenient. For instance, the
heat current $\sim \partial T(r_m)/ \partial r_i$ at the
real space point $r_m$ now presumes the knowledge (not
usually available) of the global transformation,
$r_m\leftrightarrow a_m$, as $\partial T(r_m)/\partial
r_i = [\partial T(a_m)/\partial a_k](\partial
a_k/\partial r_i)$. Similarly, with $\mbox{\boldmath$g$}$
the momentum density, the angular momentum density is
${\bf r} (a_m)\times \mbox{\boldmath$g$}(a_m)$ rather
than $\mbox{\boldmath$a$}
\times\mbox{\boldmath$g$}(a_m)$. (If the system is only
weakly deformed, with $u_i=r_i-a_i$ small, the above
differences between $r_i$ and $a_i$ may be neglected to
linear order.) Finally, more specific to the issue at
hand, our equations need to contain both the elasticity
theory and the liquid hydrodynamics. The latter, however,
is usually and concisely given in the Euler notation.

The second, and the only actually viable,
choice is to take all variables including the strain
tensor in the local, Eulerian notation, as functions
of $r_m$.
We shall
therefore employ the Eulerian strain tensor \cite{Lub2,Temmen},
introduced via ${\rm d}r^2_i-{\rm
d}a^2_i(r_m)=2U_{ik}(r_m){\rm d}r_i{\rm d}r_k$,
where $U_{ik}(r_m)= \textstyle\frac{1}{2}[\partial
u_i/\partial r_k+\partial u_k/\partial r_i-
(\partial u_j/\partial r_i)(\partial u_j/\partial
r_k)]$ and $u_i(r_m)=r_i-a_i(r_m)$.

In this context, there is a second, somewhat subtle
point: We need to eliminate the displacement field $u_i$,
and deal exclusively with the initial coordinate $a_i(r)$
and the strain $U_{ik}$ in the elasticity theory. This is
possible because starting again from ${\rm d}r^2_i-{\rm
d}a^2_i(r_m)=2U_{ik}(r_m){\rm d}r_i{\rm d}r_k$, we find
that the strain tensor may be written as
\begin{equation}\label{one}
U_{ik}=\textstyle\frac{1}{2}[\delta_{ik}-(\partial
a_\alpha/\partial r_k) (\partial a_\alpha/\partial
r_i)],
\end{equation}
with no need whatever for a detour via $u_i$.
This is necessary because the introduction of $u_i$
destroys a symmetry and represents an arbitrary (gauge)
choice. As discussed, $a_i$ and $r_i$ are vectors of
different spaces, so they transform as vectors under
rotations in initial and real space, respectively. The
introduction of the displacement fixes both spaces with
respect to each other, and prohibits the rotation of
either space alone. On the other hand, the elastic energy
is independent of the orientation of the initial space,
the fictitious unstressed body. Given any transformation
$a_m\leftrightarrow r_m$, we should still be free to take
a global but arbitrary rotation of all $a_i$, ie, rotate
the initial space with respect to the real space.
Therefore, $a_\alpha$ and $r_i$ are indeed vectors of two
different spaces, and a quantity such as $\nabla_{k}
a_\alpha\equiv\partial a_\alpha/\partial r_i$ is a vector
both in real and initial space, (a bi-vector,) and not a
second rank tensor.  We use Latin and Greek indices
to denote the components (x,y,z) in real space, and
(1,2,3) in initial space, respectively. (Clearly, this
renders the fact that the displacement $r_i-a_\alpha$ is
an oxymoron rather obvious.)

We now proceed to understand that the bi-vector $\nabla_i
a_\alpha$ not only contains the information about the
strain, as shown in Eq~(\ref{one}), but also that about
the local orientation. The {\em polar decomposition
theorem} (cf W. Noll, p.65 ff, Vol.2 of~\cite{truesdell})
states
\begin{equation}\label{two}
\nabla_i a_\alpha= R_{\alpha j} \Xi_{ij},
\end{equation}
where $R_{\alpha j}$ is the rotation matrix that rotates
the local preferred directions in real space back to the
global ones in initial space, while $\Xi_{ij}$ is a
symmetric matrix that deviates from $\delta_{ij}$ only
for finite strains. Consider first the unstrained case
$\Xi_{ij}=\delta_{ij}$: Because of ${\rm
d}a_\alpha=(\nabla_{i} a_\alpha){\rm d}r_i$ with ${\rm
d}a^2_\alpha={\rm d}r^2_i$, the gradient $\nabla_i
a_\alpha$ is indeed a rotation matrix $R_{\alpha j}$, and
must satisfy $R_{\alpha j}R_{\alpha k}=\delta_{jk}$,
$R_{\alpha j}R_{\beta j}=\delta_{\alpha\beta}$. For
finite strains, Eq.~(\ref{one}) implies $\delta_{ij}
-2U_{ij}= R_{\alpha  k}\Xi_{ik}\, R_{\alpha l} \Xi_{lj} =
\Xi_{ik}\Xi_{kj}$, the square root of which is
\begin{equation}\label{PDT}
\Xi_{ij}=\sqrt{\delta_{ij}-2U_{ij}} \approx (
\delta_{ij}-U_{ij}-\textstyle{\frac{1}{2}} U_{ik}
U_{kj}\cdots).
\end{equation}
[This expansion is valid for small strains $U_{ij}$, but
arbitrary rotations $R_{\alpha j}$. The square root of a
matrix is defined by its power series. One can verify
Eq~(\ref{PDT}) by calculating $\Xi_{ik}\Xi_{kj}$.]

In accounting for solid behaviour, we need to keep
track of the local preferred directions, or
$R_{\alpha i}$, which may vary considerably by
accumulation over a long distance, even if the
strain is small -- think of a sheet of single
crystal, slightly bent over a long stretch to form a
tube of large radius. Let us consider as an example
the harmonic approximation for the energy $E = \int
\epsilon \,\, {\rm d}V$,
\begin{eqnarray}\label{ep1}
\epsilon=\frac{1}{2} K_{ijkm}U_{ij}U_{km}=
\textstyle\frac{1}{2} K_{\alpha\beta\gamma\delta}
U_{\alpha\beta}U_{\gamma\delta} \\  U_{ij}= R_{\alpha
i} R_{\beta j} U_{\alpha\beta},\,\,\, K_{ikjm}=
R_{\alpha i} R_{\beta j} R_{\gamma k} R_{\delta m}
K_{\alpha\beta\gamma\delta} \label{ep2}
\end{eqnarray}
where $U_{\alpha\beta}$ and
$K_{\alpha\beta\gamma\delta}$ are the attendant
quantities in the initial space. ($\epsilon$ has the
same form in both spaces because $R_{\alpha i}$
annihilates pairwise.) The elements of
$K_{\alpha\beta \gamma\delta}$ are constant. A cubic
crystal for instance has three independent elements,
of which one is given as
$K_{1111}=K_{2222}=K_{3333}$, implying that the
compressional energy is the same along the three
initial space directions 1, 2, and 3. This is not
the case for a hexagonal crystal, for which the
compressional energy along 3 is different from 1 or
2. The real space matrix $K_{ijkm}$ depends on
$R_{\alpha i}$ and varies in space, as Eq(\ref{ep2})
shows, because the symmetry axis 3 (of a hexagonal
crystal) may in real space point in any direction,
and vary spatially. So $\epsilon$ is a function of
$U_{ij}$ and $R_{\alpha i}$. Writing ${\rm
d}\epsilon=\Psi_{ij}{\rm d} U_{ij} + \chi_{\alpha
i}{\rm d}R_{\alpha i}$, the conjugate variables
$\Psi_{ij}=K_{ijkm}U_{km}$ and $\chi_{\alpha i}=
2R_{\beta j} R_{\gamma k} R_{\delta m}$ $
K_{\alpha\beta\gamma\delta} U_{ij}U_{km}$ are given
by differentiating Eqs~(\ref{ep1}) and (\ref{ep2}).
The energy $\epsilon$ depends on $U_{ij}$ and
$R_{\alpha i}$ beyond the validity of
Eq~(\ref{ep1}), so ${\rm d}\epsilon=\Psi_{ij}{\rm d}
U_{ij} + \chi_{\alpha i}{\rm d}R_{\alpha i}$ is
general\-ly valid -- though the explicit form of
$\Psi_{ij}$ and $\chi_{\alpha i}$ will vary.

The term $\chi_{\alpha i} d R_{\alpha i}$ is rarely
included in the energy for solids~\cite{llelastic},
which renders the resultant formulas valid only for
small deformations or isotropic solids. For the
latter systems there is no preferred direction to
keep track of locally, so $K_{ijkm}$ will not depend
on $R_{\alpha i}$, and we can set $\chi_{\alpha
i}=0$. This is easiest seen
in the harmonic approximation, Eq~(\ref{ep1}), where
$K_{\alpha\beta\gamma\delta}=
(K_L-K_T/3)\delta_{\alpha\beta}
\delta_{\gamma\delta} + K_T/2(\delta_{\alpha\gamma}
\delta_{\beta\delta}+ \delta_{\alpha\delta}
\delta_{\beta\gamma})$ due to isotropy. Inserting
this into Eq~(\ref{ep2}), we again obtain
$K_{ijkm}=(K_L-K_T/3)\delta_{ij} \delta_{km} +
 K_T/2(\delta_{ik} \delta_{jm}+ \delta_{im}
\delta_{jk})$, manifestly independent of $R_{\alpha
i}$.

Returning to anisotropic systems, the 9 variables of
$\nabla_i a_\alpha$ are equivalent to the 3 of
$R_{\alpha i}$ and the 6 of $U_{ij}$ (or
$\Xi_{ij}$), see Eqs.~(\ref{two}) and (\ref{PDT}).
So we can conveniently write $\Psi_{ij}{\rm d}U_{ij}
+ \chi_{\alpha i} {\rm d}R_{\alpha i}=\psi_{\alpha
i}{\rm d}\nabla_i a_{\alpha}$ where $\psi_{\alpha
i}=\Psi_{km}({\partial U_{km}}/{\partial \nabla_i
a_{\alpha}}) + \chi_{\beta k}({\partial R_{\beta
k}}/{\partial\nabla_i a_{\alpha}})$. Under a real
space rotation of the angle ${\rm d}\theta_i$,
scalars are invariant, ${\rm d}\epsilon=
\psi_{\alpha i} {\rm d}\nabla_i a_{\alpha} =0$, but
vectors and tensors are not, ${\rm d}\nabla_i
a_{\alpha}=\epsilon_{ijk}\nabla_j a_{\alpha}{\rm
d}\theta_k$, so $\psi_{\alpha
i}\epsilon_{ijk}\nabla_j a_{\alpha}=0$, or
$(\psi_{\alpha j} \nabla_{i}
a_{\alpha})=(i\leftrightarrow j)$. Similarly, ${\rm
d}R_{\alpha i}=\epsilon_{ijk}R_{\alpha j}{\rm
d}\theta_k$, ${\rm d}U_{im}=\epsilon_{ijk}U_{jm}{\rm
d}\theta_k+\epsilon_{mjk}U_{ij}{\rm d}\theta_k$, so
$(\Psi_{ik}U_{jk}+ \Psi_{ki}U_{kj}+ \chi_{\alpha
i}R_{\alpha j})=(i\leftrightarrow j)$. (These
constraints on $\psi_{\alpha j}$, $\Psi_{ki}$, and
$\chi_{\alpha i}$ can be used to show the symmetry
of the stress tensor $\sigma_{ij}$ below.)

More generally, $\epsilon$ also depends on the mass,
entropy and momentum density, $\rho$, $s$, and $g_i$. So
the final thermodynamic expression for an elastic medium
is
\begin{equation}
\label{Gibbs} {\rm d}\epsilon = T{\rm d}s + \mu {\rm
d}\rho + v_{i}{\rm d}g_{i} + \psi_{\alpha i}{\rm d}
\nabla_i a_{\alpha}.
\end{equation}

Turning now to dynamics, the equation of motion for
$a_{\alpha}$ is
\begin{equation}\label{adyn}
\textstyle\frac{\rm d}{{\rm d}t}a_{\alpha}
\equiv{\dot a_{\alpha}+ v_{k} \nabla_{k}
a_{\alpha}=-Y_\alpha}.
\end{equation}
In equilibrium, with the dissipative contribution
$Y_\alpha$ vanishing, this equation simply states the
fact that the initial coordinate $a_{\alpha}$ of a mass
point does not change when one moves with it.

The entropy production $\dot s+\nabla_i(sv_i- f_i)=R/T$,
conservation of mass and momentum, $\dot
\rho+\nabla_i(\rho v_i) =0$, $\dot g_{i} +
\nabla_{j}(\sigma_{ij}-\sigma^{\rm D}_{ij})=0$, and
Eq.~(\ref{adyn}) represent the complete hydrodynamic
theory of solids, where
\begin{eqnarray}
\label{stress1} \sigma_{ij} = p\delta_{ij} + v_{i}
g_{j} + \psi_{\alpha j} \nabla_{i} a_{\alpha},\\
R=f_i\nabla_iT+\sigma^{\rm D}_{ij}A_{ij} -Y_{\alpha}
\nabla_k\psi_{\alpha k}, \label{R}\end{eqnarray}
[with $A_{ik}\equiv{\textstyle \frac{1}{2}}
(\nabla_i v_k+\nabla_k v_i)$, $p\equiv-\epsilon+Ts+
\mu\rho+ v_ig_i$] are unambiguously given by
thermodynamics, Eq.~(\ref{Gibbs}), via the
hydrodynamic standard procedure. Eq.~(\ref{R})
implies that the three fluxes $f_i, \sigma^{\rm
D}_{ij}, Y_{\alpha}$ are linear combinations of the
three forces $\nabla_iT, A_{ij},
\nabla_k\psi_{\alpha k}$. These give rise,
respectively, to the dissipative phenomena of
diffusive heat current, viscous stress, and defect
diffusion. The structure of the linear combination,
ie the symmetry of the Onsager coefficients, are
given by the symmetry group of the
crystal~\cite{llelastic}. For isotropic solids, we
have only diagonal terms, especially
$f_i\sim\nabla_iT$ and $Y_{\alpha}\sim
\nabla_k\psi_{\alpha k}$. As discussed at length in
\cite{mpp,cohen}, it is incorrect to take the latter
contribution as zero: The initial coordinate of a
mass point may change, $\dot a_{\alpha}\not= 0$, in
the absence of any mass current, $v_i=0$, when there
is diffusive motion of vacancies. Conversely,
motions of interstitials involve mass current,
$v_i\not=0$, but no change of crystal points, $\dot
a_\alpha=0$.

Since $\nabla_ia_\alpha$ as a variable is completely
equivalent to $U_{ij}$ and $R_{\alpha i}$, the
equation of motion~(\ref{adyn}) for $\dot a_\alpha$
may always be rewritten as two equations of motion,
for $\dot U_{ij}$ and $\dot R_{\alpha i}$. Though
rather more complicated, this is certainly closer to
the conventional elasticity theory. With the help of
Eqs.~(\ref{two}) and (\ref{PDT}), we rewrite
Eq.~(\ref{adyn}) as
\begin{eqnarray}
 \label{12}
  2\textstyle{\frac{\rm d}{\rm dt}}
U_{ij} =  [\Xi_{jl} \Xi_{lk} \nabla_{i} v_{k} +
R_{\alpha k} \Xi_{jk} \nabla_{i} Y_{\alpha}] + [ i
\leftrightarrow j],
\\
\label{11} \Xi_{ij} R_{\alpha l}\textstyle{\frac{\rm
d}{\rm dt}} R_{\alpha j} = - \Xi_{lj} \nabla_{i}
v_{j} - R_{\alpha l} \nabla_{i} Y_{\alpha} -
\textstyle{\frac{\rm d}{\rm dt}}\Xi_{il},
\end{eqnarray}
which may be approximated by taking ${U_{ij}}$, $A_{ij}$, $Y_{\alpha}$ and
$R_{\alpha j}\frac{\rm d}{{\rm d}t}{R_{\alpha i}} - \omega_{ij}$
 as small quantities (with
$2\omega_{ij} \equiv
\nabla_{j} v_{i} - \nabla_{i} v_{j}$).
To second order in the small
quantities, though neglecting terms of order $U_{kj}
\nabla_iY_\alpha$, the result  is
\begin{eqnarray}
\label{dotU}\textstyle{\frac{\rm d}{\rm dt}} U_{ij} -
A_{ij} =[\textstyle\frac{1}{2} (\nabla_iY_\alpha)
R_{\alpha j}-(\nabla_iv_k) U_{kj}] + [ i \! \leftrightarrow
\! j]
\\
\label{dotO} R_{\alpha j}\textstyle{\frac{\rm d}{\rm dt}}
R_{\alpha i} -  \omega_{ij} =[\textstyle{ \frac{1}{2}
R_{\alpha i} \nabla_{j} Y_{\alpha} + \frac{1}{2}} U_{jk}
A_{ik}] - [ i \!\leftrightarrow \!j]
\end{eqnarray}
Written in the conjugate variables of $U_{ij}$ and
$R_{\alpha i}$, the stress tensor Eq.~(\ref{stress1})
reads
\begin{eqnarray}
\label{stress} \sigma_{ij} = p\delta_{ij} + v_{i}
g_{j} - \Psi_{ij}+ \Psi_{ki} U_{jk}
+ \Psi_{kj} U_{ik} \nonumber\\+ \chi_{\alpha j}R_{\alpha i}+
\textstyle{1\over2}\chi_{\alpha k}( U_{ki} R_{\alpha
j} + U_{kj} R_{\alpha i}).
\end{eqnarray}

This ends the presentation of the hydrodynamic
theory of solids. The noteworthy point is: The
derivation is completely cogent, as not a single
step in it is discretionary; hence the above set of
differential equations, given between
Eq~(\ref{Gibbs}) and (\ref{R}), is unique -- any
other theory is either algebraically equivalent, or
wrong. Conversely, these equations account for any
solid system, including crystals of all symmetry
groups and glasses. This pertains especially to the
nonlinear structure, important if one is to account
for large displacement and rotation, strong
compression and shear. These are usually  small in
bulk crystals, but quite large in complex liquids.
In awareness of this, many nonlinear models for
convective-like nonlinearities have been suggested
\cite{truesdell,noll}, though none was constructed
to contain the nonlinear solid limit.

To generalize our results to visco-elastic Non-Newtonian
fluids, we note first that the solid hydrodynamics
contains the liquid hydrodynamics, and one can reduce the
former to the latter by setting to zero the elastic
stress $\psi_{\alpha i}=\partial \epsilon /\partial
(\nabla_i a_\alpha)$. Confining ourselves to isotropic
systems, it suffices to set $\Psi_{ij}=0$, because
$\chi_{\alpha i} =0$ already holds. Taking $U_{ij}$ as
the variable that relaxes as long as $\Psi_{ij} \neq 0$,
we connect the isotropic solid dynamics to fluid dynamics
such that the former holds in the high frequency regime
(where the relaxation is negligible) and the latter in
the low frequency regime (where relaxation is dominant).
Therefore, we proceed by allowing a relaxation term
$X_{ij}$ in Eq.~(\ref{dotU}), $\dot
U_{ij}+\cdots=X_{ij}$. It leads to an additional term in
the entropy production, Eq~(\ref{R}),
$R=\cdots-X_{ij}\Psi_{ij}$, which implies
$X=-\sum\alpha_i\Psi^i$ in an expansion, or to lowest
order, $X^0_{ij}=-\alpha_T \Psi^0_{ij}$ and
$X_{kk}=-\alpha_L\Psi_{ll}$. (The superscript $^0$
denotes the traceless part of the given tensor.) So we
have
\begin{eqnarray}
\textstyle\frac{\rm d}{{\rm d}t}U_{ij} &-& A_{ij}+
[(\nabla_iv_k) U_{kj}-\textstyle\frac{1}{2}
(\nabla_iY_\alpha) R_{\alpha j}+i\leftrightarrow j ]
\nonumber\\ &=&
-\alpha_T\Psi^0_{ij}-\alpha_L\Psi_{kk}\delta_{ij}/3,
\label{relax}
\end{eqnarray}
with $\alpha_T,\alpha_L$ denoting two transport
coefficients. To understand the added terms, one can
use the example of the harmonic approximation,
Eq~(\ref{ep1}), yielding $\alpha_T\Psi^0_{ij}=
\alpha_TK_TU^0_{ij}= U^0_{ij}/\tau_T$ and
$\alpha_L\Psi_{kk}= \alpha_LK_LU_{kk}=
U_{kk}/\tau_L$. Clearly, this implies relaxation for
$U^0_{ij}$ and $U_{kk}$, with the respective
relaxation times $\tau_T$ and $\tau_L$. (In
principle, there are two thermodynamic cross
derivatives, $\delta\Psi_{kk}=
K_\rho\delta\rho+K_T\delta T$.)

Note the universality of the results, especially the
convective terms $\sim(\nabla_iv_k)$, which remarkably
are not preceded by any material-dependent coefficients.
Their form is independent from the above approximation
for $X_{ij}$ and will remain unchanged even if additional
variables are introduced, eg to account for the
material-dependent rheological behavior such as shear
thinning and normal stress differences.

To account for large deformation, rotation and
velocity, many different nonlinearities, as
mentioned, have been suggested and implemented in
the rheology literature, of which the two more
popular ones are the upper and lower convective
derivatives. Denoting an arbitrary matrix as
$(\mbox{\boldmath$*$})$, the former is defined as
$\hat D_u(\mbox{\boldmath$*$})\equiv
(\partial/\partial t +\mbox{\boldmath$v\cdot
\nabla$})(\mbox{\boldmath$*$}) \, + \,
(\mbox{\boldmath$\nabla v$}) (\mbox{\boldmath$*$})
\, + \, (\mbox{\boldmath$*$})
(\mbox{\boldmath$\nabla v$})^{T}$, the latter as
$\hat D_\ell(\mbox{\boldmath$*$})\equiv(
\partial /\partial t+\mbox{\boldmath$v\cdot\nabla$})
(\mbox{\boldmath$*$})\,-\,(\mbox{\boldmath$\nabla v$})^T
(\mbox{\boldmath$*$})\,-\, (\mbox{\boldmath$*$})
(\mbox{\boldmath$\nabla v$})$. Both are derived by
invoking some variant of a postulated general principle,
usually referred to as the ``material frame
independence". In the rheology literature
\cite{truesdell,noll}, $(\mbox{\boldmath$*$})$ is the
stress tensor, taken as independent, but in principle it
could also be the strain tensor.

Reviewing the many equations of motion considered
above, it is easy to see that Eqs.~(\ref{dotU}) and
(\ref{relax}) for $U_{ij}$ can indeed be written as
$\hat D_u\,{\bf U}- {\bf A}= O({\mbox{\boldmath${
\nabla Y, \Psi}$}})$. None of the other equations
may be brought into this form: It is not valid for
$\dot a_\alpha$ and $\dot R_{\alpha i}$, see
Eqs.~(\ref{adyn}, \ref{11}, \ref{dotO}); nor for the
exact equations $\dot U_{ij}$, Eq.~(\ref{12}).
Especially, it does not hold for the stress tensor.

This is a rather serious shortcoming and subjects all
those descriptions that include upper and lower convected
derivatives, combinations thereof, or other kinds of
quadratic nonlinearities to grave doubts. To overcome
these, the authors really need to convincingly argue why
their postulated general principle overrules the simple
and physical requirement that, for infinite yield time of
the stress, the dynamics of non-Newtonian liquids such as
polymer melts is that of an isotropic elastic medium.

\end{document}